\documentclass[seceq]{ptptex}

\usepackage{graphicx}

\title{Power laws and market crashes}
\subtitle{Empirical laws on bursting bubbles}    

\author{Taisei \textsc{Kaizoji}}

\inst{Division of Social Sciences, International Christian University, \\
3-10-2 Osawa, Mitaka, Tokyo, 181-8585 Japan.}

\abst{
In this paper, we quantitatively investigate the statistical properties of a statistical ensemble of stock prices. We selected 1200 stocks traded on the Tokyo Stock Exchange, and formed a statistical ensemble of daily stock prices for each trading day in the 3-year period from January 4, 1999 to December 28, 2001, corresponding to the period of the forming of the internet bubble in Japn, and its bursting in the Japanese stock market. We found that the tail of the complementary cumulative distribution function of the ensemble of stock prices in the high value of the price is well described by a power-law distribution, $ P(S>x) \sim x^{-\alpha} $, with an exponent that moves in the range of $ 1.09 < \alpha < 1.27 $. Furthermore, we found that as the power-law exponents $ \alpha $ approached unity, the bubbles collapsed. This suggests that Zipf's law for stock prices is a sign that bubbles are going to burst.  
}

\begin{document}

\maketitle

\section{Introduction}

In the last few decades, the stock markets has been frequently visited by large market crashes. The increasingly frequent market crashes have attracted the attention of academics and policy makers. Numerous reports, commentaries, and academic books [1-8] have described the various reasons {\it why} stock market crashes have occurred\footnote{The first econophysisist to pay a lot of attention to stock market crashes was Didier Sornette [3]. Sornette, Johansen, and Bouchaud [9], and Feigenbaum and Freund [10] independently proposed a theory that large stock market crashes are analogous to critical points studied in statistical physics with log-periodic correction to scaling. Their theory predicts the existence of a log-frequency shift over time in the {\it log-periodic oscillations} prior to a crash. Lillo and Mantegna [11,12] studied the relaxation dynamics of a stock market just after the occurrence of a large crash by investigating the number of times the absolute value of an index return exceeded a given threshold value. They show that the empirical observation of a power law evolution of the number of events exceeding the selected threshold (a behavior known as the {\it Omori law} in geophysics) is consistent with the simultaneous occurrence of (i) a return probability density function characterized by a power law asymptotic behavior and (ii) a power law relaxation decay of its typical scale.}.  One of the greatest myths is that market crashes are random, unpredictable events. Can large market crashes be forecast? Do common warning signs exist which characterize the end of a bubble and the start of a market crash? \par
My previous work [13] presents an interacting-agent model of speculative activity explaining bubbles and crashes in stock markets. The model describes stock markets through an infinite-range Ising model to formulate the tendency of traders getting influenced by the investment attitude of other traders, and demonstrates that a burst of speculative bubbles is considered qualitatively and quantitatively as the phase transition from a bull market phase to a bear market phase.\par
More recently, Kaizoji and Kaizoji [14] quantitatively investigated the statistical properties of an ensemble of land prices in Japan in the 22-year period from 1981 to 2002, corresponding to the period of Japan's bubble economy and the crashes, to study warning phenomenon of large market crashes. We found that the tail of the complementary cumulative distribution function of the ensemble of land prices in the high price range is well described by a power-law distribution, $ P(S>x) \sim x^{-\alpha} $, and furthermore, that as the power-law exponents $ \alpha $ approached unity, the bubbles collapsed. We reaffirmed our findings in [14]. We abstracted the land prices in the Tokyo area from a database of the {\it assessed value of land} that is made public once a year by the Ministry of Land, Infrastructure and Transport Government of Japan, and investigated the statistical properties of an ensemble of land prices for each of the years in the 22-year period from 1981 to 2002, corresponding to the period of the Japan's bubble economy and the crashes. Figure 1(a) plots the cumulative probability distribution of the ensembles of land prices in the four years 1985, 1987, 1991, and 1998, on the log-log scale. We found that the tail of the complementary cumulative distribution function of the ensemble of land prices in the high price range is described by a power-law distribution, $ P(S>x) \sim x^{-\alpha} $. We use ordinary least squares (OLS) regression in log-log coordinates in order to determine the exponent $\alpha$ for the data of land prices for each of the 22 years from 1981 - 2002. Figure 1(b) indicates the movement of the exponent $ \alpha$. The exponent $\alpha$ continued to decrease toward unity between 1981 and 1987, and during the period of peaks and bursts of bubbles from 1987 to 1992, the exponent $\alpha$ hovered around unity. After the crashes, the exponent $\alpha$ continued to increase between 1992 and 2001. This finding suggests that the threshold value of the exponent $\alpha$ that cause bubbles to burst is unity\footnote{Yamano [15] investigated the complementary cumulative distribution of the absolute value of year-on-year percentage changes of land prices in Japan. His work has been motivated by  generalized thermostatistics. Andersson, et. al. [16] proposed a network model of urban economy that assumed that the market dynamics that generate land values can be represented as a growing scale-free network.}. \par
The aim of this paper is to attempt to extend the observation into stock markets, and to examine if the same as phenomena we found in bursts of the land markets can be observed in stock market crashes. To do so, we focused our attention on the statistical properties of ensembles of stock prices on Japan's internet bubble (or 
the dot-com bubble), ending with the bursts in 2000. Using the $1200$ stocks listed and traded on the Tokyo Stock Exchange for over 20 years, we formed ensembles of the daily stock prices in the 3-year period from January 4, 1999, to December 28, 2001, which corresponds to the period of Japan's internet bubble and its' crash. We found that the tail of the complementary cumulative distribution function of the ensemble of stock prices in the high value range is well described by a power-law distribution, $ P(S>x) \sim x^{-\alpha} $, with an exponent that moves in the range of $ 1.09 < \alpha < 1.27 $. Furthermore, we found that as the power-law exponents $ \alpha $ approached unity, the internet bubbles collapsed. 
This suggests that Zipf's law for an ensemble of the stock prices is a sign that bubbles will burst. 

\begin{figure}
\begin{center}
  \includegraphics[height=14cm,width=13cm]{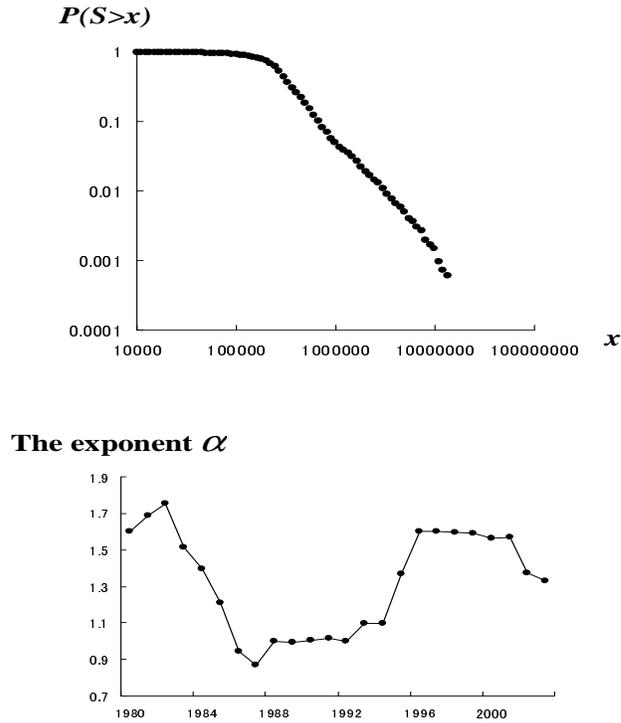}
\end{center}
\caption{(a) The Log-log plot of the complementary cumulative distribution $ P(S > 
x) $ of the land prices' ensembles of the Tokyo area in 1997.  
(b) The movement of the power-law exponent $ \alpha $ in the 22-year period 
from 1981 to 2002.}
\label{fig1}
\end{figure}

\section{The internet bubble}

In the 1990s, the personal computer, software, telecommunications and internet were 
rapidly gaining acceptance for business and personal use. Computer-related 
technology drove the powerful bull market trend of global markets. Several 
economists even postulated that we were in a {\it New Economy}, where inflation was 
nonexistent and stock market crashes were obsolete. Investors over the world were 
euphoric and believed in the fallacy of a perpetual bull market. Large scale stock 
speculation occurred, causing a worldwide mania. By the late 1990s, many technology 
companies were selling stock in initial public offerings (IPOs). Most initial 
shareholders, including employees, became millionaires overnight. By early 2000, 
reality began to sink in. Investors soon realized that the dot-com dream was really 
a bubble. Within months, the Nasdaq crashed from 5,000 to 2,000. In the Japanese stock 
market, many {\it high-tech} stocks which were worth billions were off the map as 
quickly as they appeared, too. Panic selling ensued as investors lost trillions of dollars. Numerous accounting scandals came to light, showing how many companies had artificially inflated earnings. A lot of shareholders were crippled. Once again, we saw the development of a bubble and the inevitable stock market crash that always follows it. \par
Figures 2(a) and 2(b) show the movement of the mean and variance of the stock prices' ensemble distribution for each day of the three-year period from January 4, 1999 to December 28, 2001. Bursting internet bubbles is explicit in the movement of the variance rather than the movement of the mean. The variance of the stock prices abnormally surged until the middle of 2000. The variance peaked on August 17, 2000, 
and dropped on December 28, 2001 by 93.3 percent. The reason why the abnormal enlargement of the variance of stock prices occurred is probably because capital investment was extremely concentrated on stocks of industries which were related to the personal computer, software, telecommunications and internet, so that the prices of a tiny number of the large, high-tech stocks surged from the beginning of 2000. 

\begin{figure}
\begin{center}
  \includegraphics[height=14cm,width=12cm]{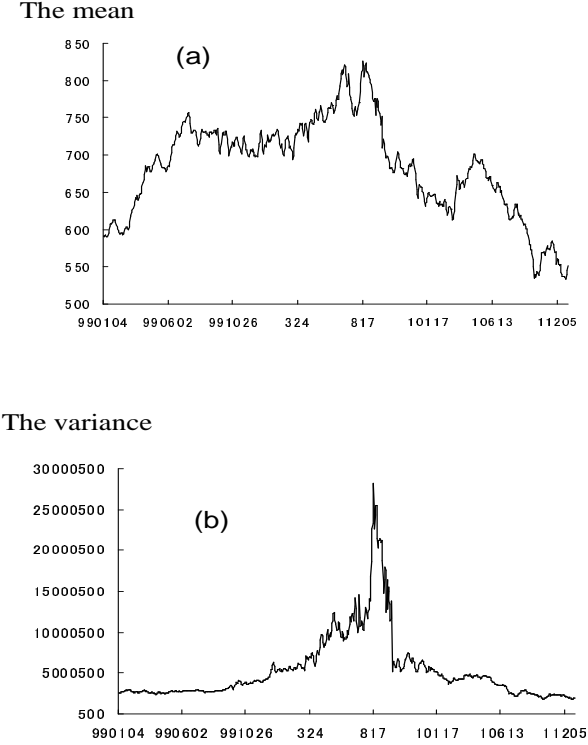}
\end{center}
\caption{(a) The Log-log plot of the complementary cumulative distribution $ P(S > 
x) $ of ensembles of the land prices of the Tokyo area in 1997.  
(b) The movement of the power-law exponent $ \alpha $ in the 22-year period 
from 1981 to 2002.}
\label{fig2}
\end{figure}

\section{Power-law for ensembles of stock prices}

Figure 3(a) shows the complementary cumulative distribution of the stock prices' ensemble of 1,200 companies listed on the Tokyo Stock Exchange for each of the ten days selected randomly on the log-log scale. We found that the tail of the complementary cumulative distribution function of the stock prices' ensemble in the high price range is well described by a power-law distribution, $ P(S>x) \sim x^{-\alpha} $. To confirm the robustness of the above analysis, we repeated this analysis for each of the trading days in the 3-year period from January 4, 1999, to December 28, 2001, which correspond to the period of Japan's internet bubble and the collapse. Figure 3(b) shows the complementary cumulative distribution of the stock prices' ensemble drawn by using a whole database of stock prices of 1200 companies in the 3-year period from January 4, 1999 to December 28, 2001. The tail of the distribution in the high prices is also described by a power-law decay, $ P(S > x) \sim x^{-\alpha} $. \par

\begin{figure}
\begin{center}
  \includegraphics[height=14cm,width=13cm]{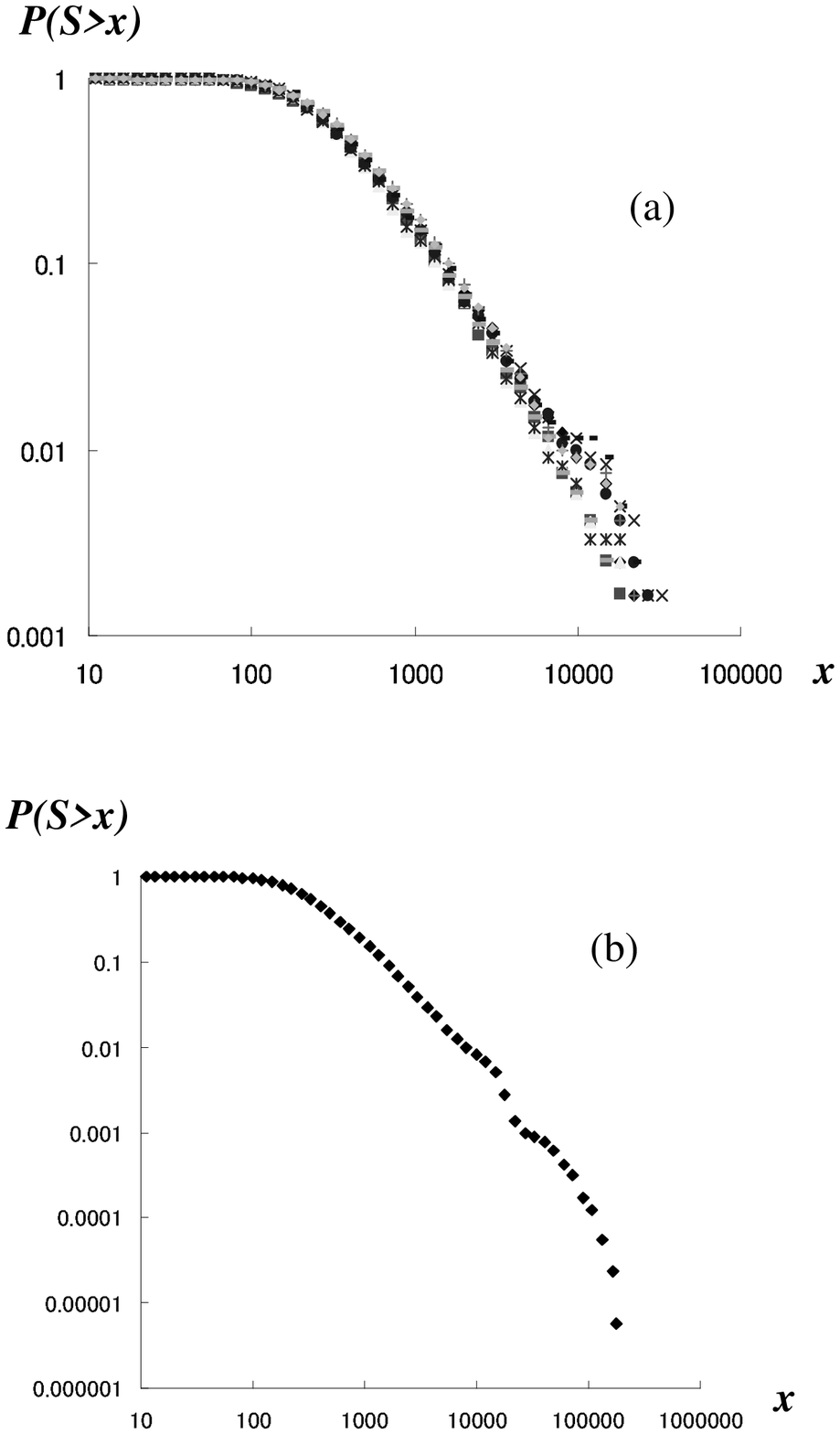}
\end{center}
\caption{(a) The Log-log plot of the complementary cumulative distribution $ P(S > 
x) $ of ensembles of the 1200 stock prices for each of the ten days selected 
randomly. (b) The complementary cumulative distribution of stock price 
ensemble drawn by using a whole database of stock prices of 1200 companies in the 
3-year period from January 4, 1999 to December 28, 2001.}
\label{fig3}
\end{figure}

Figure 4(a) shows the movement of the power-law exponent $ \alpha $ in the period from January 4, 1989, to December 30, 1992. We used ordinary least squares (OLS) regression in the log-log coordinates in order to determine the exponent $\alpha$. The power-law exponent $ \alpha $ moved in the range of $1.08 < \alpha < 1.27 $. The exponent $\alpha$ continued to decrease toward unity during 1999, and the exponent $\alpha$ hovered around unity during the period of the peak of bubbles in 2000. The exponent $\alpha$ started to increase from the middle of 2000, when the bubbles crashed. \par
These results lead us to the conclusion that the power-law distribution of stock prices' ensemble is very strong, but the temporary ensemble distribution changes over time. This means that the stochastic process on prices is non-stationary. Furthermore, these findings suggest that the threshold value of the exponent $\alpha$ that cause bubbles to burst is unity. We interpreted this as follows. The mean of the power-law distribution is equal to $ 1/(\alpha - 1) $. Therefore, as the power-law exponent $ \alpha $ approaches unity, the mean diverges and the bubbles inevitably collapse. From a practical viewpoint, let us also consider the implications of these empirical findings. \par
The exponent $\alpha$ of the power-law distribution can be considered a measurement of the price inequality. When the exponent $ \alpha $ is equal to unity, the price inequality in the stock market reaches a maximum. We can demonstrate this using the Gini coefficient $(G)$ [17], known as the index for wealth concentration. Theoretically, the Gini coefficient ranges from zero, when all land areas are equal in price, to unity, when one land area has the highest price and the rest none. If the exponent $ \alpha $ is close to unity, it means that the Gini coefficient will be close to unity. On the other hand, the Gini coefficient can be written as $G=1/(2 \alpha - 1) $. Therefore, when the power-law exponent $ \alpha $ is equal to unity, the Gini coefficient is also equal to unity. Figure 4(b) shows the movement of the Gini coefficient of the stock price ensemble. The inequality of the ensemble of stock prices measured by the Gini coefficient increased drastically from 0.58 on January 4, 1999 to 0.67 on August 18, 2000, and particularly during the final term of the internet bubble, the Gini coefficient was nearly 70 percent, and price inequality reached the breaking point. 

\begin{figure}
\begin{center}
  \includegraphics[height=14cm,width=13cm]{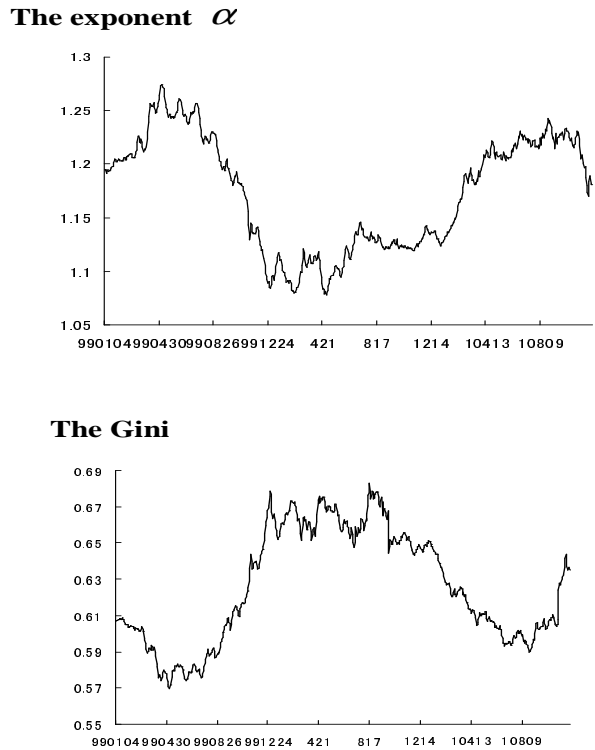}
\end{center}
\caption{(a) The movement of the power-law exponent $ \alpha $ in the 3-year period 
from January 4, 1999 to December 28, 2001. (b) The movement of the Gini coefficient 
of the stock prices $ \bar{S} $ in the 3 year period from January 4, 1999 to December 
28, 2001.
}
\label{fig4}
\end{figure}

\section{Concluding remarks}
In this paper, we focused our attention on the statistical properties of ensembles of 
stock prices. Using $1200$ stocks traded on the Tokyo Stock Exchange, we formed 
ensembles of daily stock prices in the period of Japan's internet bubble. 
We found that the tail of the complementary cumulative distribution function of the 
ensemble in the high price range is well described by a power-law distribution, 
$ P(S>x) \sim x^{-\alpha} $, and when the power-law exponent $ \alpha $ approaches 
unity, the bubble collapses. These empirical findings are the same as those on the 
Japan's land market crashes in the early 1990s. The internet bubble in 2000 is defined 
as an extraordinary enlargement of the inequality of stock prices as well as Japan's land market crash in the early 1990s. 
Our findings suggest that (i) bubbles are caused by an over-concentration of investment 
capital, and (ii) the stochastic dynamics of asset prices is non-stationary, and that 
(iii) Zipf's law for stock prices is an indication that bubbles will burst. The next step is to model these findings. A theoretical study will be conducted in the future. 

\section{Acknowledgement}
My special thanks are due to the members of financial engineering 
Group in Nikkei Media Marketing Inc, and Ms. Michiyo Kaizoji for their assistance 
in providing and analyzing the data. I also wish to thank the organizers of CN-Kyoto 
for giving me a precious occasion to speak about my research, and participants of 
CN-Kyoto for valuable comments and helpful suggestions. Financial support by the 
Japan Society for the 
Promotion of Science under the Grant-in-Aid, No. 15201038 is gratefully acknowledged. 
All remaining errors, of course, are ours. 

%

\end{document}